\documentclass[a4paper,prb,showpacs,superscriptaddress]{revtex4}

\usepackage{epsfig}

\newcommand{\beq}{\begin{equation}}
\newcommand{\eeq}{\end{equation}}
\newcommand{\beqa}{\begin{eqnarray}}
\newcommand{\eeqa}{\end{eqnarray}}

\begin{document}


\thispagestyle{empty}
\vskip 3cm

\title{Study of ${\rm CdI_2}$ nanocrystals dispersed in amorphous ${\rm
Sb_2S_3}$ matrix.} 
\author{P. Arun}
\affiliation{Department of Physics \& Electronics, S.G.T.B. Khalsa College, 
University of Delhi, Delhi - 110 007, India}
 \email{arunp92@physics.du.ac.in}

\begin{abstract}
Crystalline nanoparticles of cadmium iodide where suspended in the amorphous
matrix of antimony trisulphide. Both materials are layered structured
and have large band-gaps however cadmium iodide exhibits polytypism, i.e. it
exists in various different crystalline states. Different crystalline states
are marked by wholely different dielectric constants which give rise to
sharp surface plasmon resonance (SPR) peaks in the UV-visible. The manuscript 
details the variation in SPR's with heat-treatment of the films.  
\end{abstract}

\maketitle


\section{Introduction}

Recently, the physical properties of materials whose dimensions are highly
constraint, in nanometers, has become a matter of interest for the scientific 
community. The exponential growth in research work on this front is
testimonial of this along with their potential applications in
biosensors\cite{bio1,bio2}, sensors\cite{bio3}, medicine, hetergeneous 
catalysis\cite{chem1}, anti-reflective films\cite{opta}, solar cells/
photo-detectors and light emitting diodes\cite{led1, led2} and non-linear
optical devices\cite{opt1, opt2} etc. The subject also holds out interest to 
those in pure
sciences also. Present research works on nanoclusters are broadly in
preparing and developing processes to control the size of
nanoclusters\cite{b1,b2, b3, b4, b5, b6}, characterising them and then look for possible 
applications.

The existence of nanocrystals are manifested in the material's optical
properties. Sharp peaks are seen in the absorption spectra of the material
in the UV-visible range. The peaks are explained due to the plasmon resonance
occurring at the nanocrystal's surface. The formation of plasmon at the
surface (giving rise to what is called {\it surface plasmon resonance, SPR})
has been explained using electromagnetic theory\cite{mie}. The 
work cites the difference in the nanocrystal's dielectric constant and that
of the 
medium in which it is suspended as the cause for creations of plasmons. 
Hence, the SPR absorption depends on the dielectric properties of the 
surrounding host matrix\cite{book1}. It gives the extinction cross-section 
efficiency (which is related to the absorbance) as
\begin{eqnarray}
Q_{ext}=-{4ka \over 3}\left\{{\epsilon_2L_1(\epsilon_1-\epsilon_m)-
\epsilon_2[\epsilon_m(1-L_1)+\epsilon_1L_1] \over 2[\epsilon_m(1-L_1)+
\epsilon_1L_1]^2 + 2(\epsilon_2L_1)^2}+
{\epsilon_2(1-L_1)(\epsilon_1-\epsilon_m)- \epsilon_2[\epsilon_m(1+L_1)+
\epsilon_1(1-L_1)] \over [\epsilon_m(1+L_1)
+\epsilon_1(1-L_1)]^2+\epsilon_2^2(1-L_1)^2}\right\}\nonumber
\end{eqnarray}
\par Most of the studies available in the literature of nanocrystals 
suspended in host material, have been restricted to {\it metals (Ag and
Au)-in-glass}\cite{gold1,gold2,gold3,gold4,gold5,gold6} and {\it 
semiconductors-in-glass}\cite{semi1,semi2,semi3}. 
This makes sure that the dielectric constant of the
two materials, nanocrystals and host material, are highly diverged. However,
to the best of the author's knowledge, no example has been found where the
nanocrystals are suspended in host materials with comparable dielectric
constant. The results promises to further the understanding of 
surface plasmon formation and it's dependence on the constituent material's
dielectric constant in light of Cartoixa and Wang's\cite{prl} and similar
works. 

With this objective, samples with cadmium iodide (${\rm CdI_2}$) nanocrystals 
suspended in amorphous antimony trisulphide (${\rm Sb_2S_3}$) matrix
were fabricated by (thermal) coevaporation. The author and his 
collaborators have worked extensively with antimony
trisulphide\cite{a1,a2,a4,a5,a8} and cadmium iodide\cite{PT,PT1,PT2,PT3,rawat}. 
The properties of ${\rm Sb_2S_3}$ thin films have been 
extensively studied. The as-grown films of ${\rm Sb_2S_3}$ are invariably
amorphous in nature and only crystallize on heating at high temperatures of 
${\rm 165^oC}$\cite{thesis1}. While the electrical\cite{sb1,sb2,sb3,sb4,sb5}, 
thermo-electrical\cite{sb3} and optical properties\cite{sb6,sb7} of 
antimony trisulphide are extensively reported, till recently no
study detailed it's dielectric properties. Farid {\it et al}\cite{farid} 
report ${\rm
Sb_2S_3}$'s dielectric constant at room temperature as $\sim$10 over a large
frequency range. In contrast ${\rm CdI_2}$ films grown at room temperature
are crystalline in nature. However, depending on the growth rate, ${\rm
CdI_2}$ crystallize with different structure. This is in agreement with the
observation of large polytypism observed in ${\rm CdI_2}$\cite{tri}. 
The common and
stable structure that ${\rm CdI_2}$ crystallizes in is called 4H. For large
growth rates, ${\rm CdI_2}$ exists in nH structures where ${\rm n >4}$.
However, on heating nH structures reduce to 4H\cite{coe}. The dielectric 
constant of
${\rm CdI_2}$ films with nH structure lies between 60-190 (all possible
structures) while the dielectric constant of the stable 4H structured films
is reported to be $\sim$15-30\cite{srivastav}. This feature allows for
control of divergence in nanocrystal and host dielectric constant via growth
rate. The following sections list the experimental details and results of
the experiments.

\section{Experimental Details}

The starting material of ${\rm Sb_2S_3}$ was 99.99\% pure stoichiometric 
ingot supplied by Aldrich (USA) while ${\rm CdI_2}$ was analar grade powder, 
which was pellatized for evaporation. Samples of ${\rm CdI_2}$ and
${\rm Sb_2S_3}$ were taken in ratio of 1:10 and kept in molybdenum boat for
thermal evaporation. Thin films were grown onto microscope glass slides 
substrates at room temperature at a vacuum better than ${\rm 10^{-6}}$ Torr. 
Since ${\rm CdI_2}$ evaporates at a higher temperature than ${\rm Sb_2S_3}$,
the boats was heated to a high temperature in a short period of time. This
resulted in a fast deposition rate. The film thickness and rate of evaporation 
was monitored during it's growth by quartz crystal thickness monitor and was 
subsequently confirmed by Dektek IIA surface profiler, which uses the method of 
a mechanical stylus movement on the surface. The movement of the stylus across 
the edge of the film determines the step height or the film thickness. 
Various films of thickness between 2500 to 4000\AA\, were used during the 
experiment. However, here only results of the two extreme thickness (2500
and 4000\AA) have been report. A small piece was cut-out from the as-grown 
sample and kept aside
for characterizing experiments. The remaining portion of the slide was
heated in air (i.e. with film side facing upwards) by placing them on a copper 
block which was heated and maintained at ${\rm 150^oC}$ by a heating coil 
embedded in it. At intervals of $\sim$1hr, small pieces were cut and kept
aside for analysis. The structural, chemical composition and morphology of the 
films were determined by X-ray diffraction (Philip PW1840 X-ray diffractometer), 
photo-electron spectroscopy (Shimadzu's ESCA 750) and scanning electron 
microscope (JOEL-840) respectively. The optical properties of the films were
studied using an UV-vis spectrophotometer (Shimadzu's UV-260).

\section{Results and Discussion}
\subsection{X-Ray Diffraction studies}

As stated above, the as-grown films of ${\rm Sb_2S_3}$ are amorphous
in nature while that of ${\rm CdI_2}$ are crystalline. Thus, when grown
simultaneously one expects ${\rm CdI_2}$ crystals, present in small percentage, 
to exist embedded in the amorphous ${\rm Sb_2S_3}$ film. This is based on
the assumption no chemical reaction between the two species takes place
during evaporation. Chemical analysis of the films confirm that no reaction
took place. This can be expected since ${\rm Sb_2S_3}$ melts 
congruently\cite{3.1}. 

\begin{figure}[h]
\begin{center}
\epsfig{file=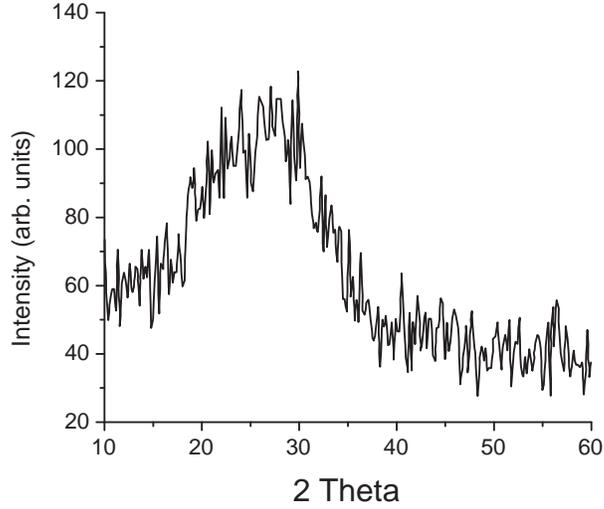,width=3.5in}
\vskip -1cm
\caption{X-ray diffractogram of sample heated for 3hrs.}
\vskip -2cm
\end{center}
\label{fig:1}
\end{figure}

X-ray diffraction studies were
done on all the samples. Without exception, all the samples, as-grown and
heated, were amorphous in nature. Figure(1) illustrates the nature of
diffractograms obtained. The small hump seen in the diffractogram is 
characteristic of the short
range ordering present in ${\rm Sb_2S_3}$ films\cite{3.8,3.9}. However, no
peaks indicating the crystalline nature of ${\rm CdI_2}$ were seen. This can
be expected since the amount of material (${\rm CdI_2}$) for scattering the
X-rays may not be sufficient. X-ray diffraction analysis hence proved futile
in estimating ${\rm CdI_2}$ nanocrystal's size. 

\subsection{Surface Morphology studies}

We have characterized all the samples for their morphology by Surface
Electron Microscopy (SEM). We have displayed few representative SEM 
micro-graphs of samples whose thickness was ${\rm 2500\AA}$ in 
fig(3). 

\begin{figure}[h]
\begin{center}
\epsfig{file=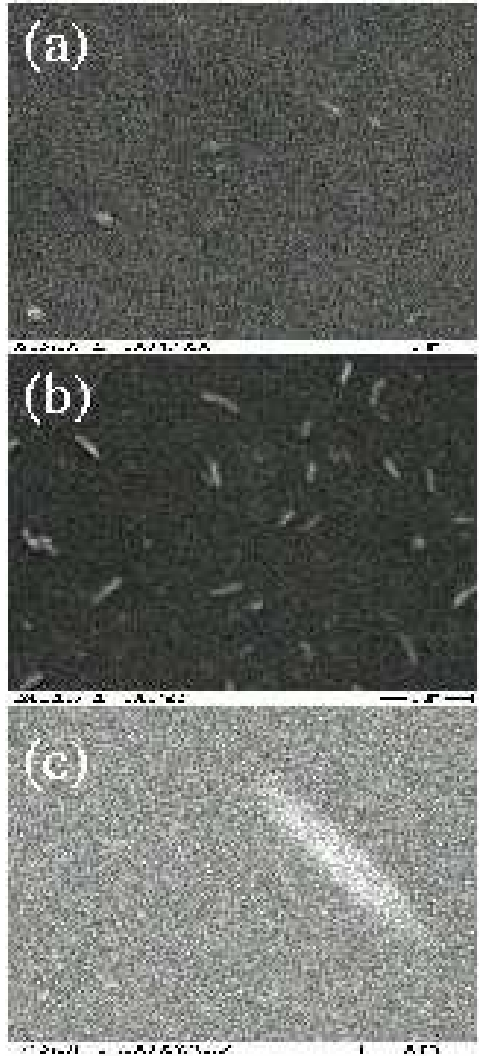,width=2.5in}
\caption{SEM micro-graphs of (a) as-grown sample of film thickness 2500\AA,\, 
(b) same sample after heating for
3hrs, and (c) a extremely magnified image of a spindle present in the last
sample.}
\vskip -2cm
\end{center}
\label{fig:1x}
\end{figure}

\par Morphological manifestations seem to be consistent with 
the structural studies, with the background of ${\rm Sb_2S_3}$ being
featureless, indicating that the material is still in it's amorphous state.
However, sparsely distributed grains of ${\rm CdI_2}$ are visible against this 
background. The possibility of grains being that of ${\rm Sb_2S_3}$ is
over-ruled since it is known to exist in amorphous state on thermal
evaporation at room temperature. 

\par The SEM micrographs of all the as-grown films showed that they 
were having grains of two distinct morphologies. 
While grains of spherical shape is predominant in the as-grown sample
(fig 2a), the second type of grains are spindle shaped. Though
they are sparsely present in the as-grown sample, their number increases with 
heating. Fig(2b) exhibits a sample heated for 3 
hours (the diffractogram of this sample is shown in fig 1). 
The simultaneous existence of both phases is evident. Further heating
leads to a further increase in the population of the spindle shaped grains. 
The grain morphologies are similar to that observed and detailed in Pankaj's
work\cite{thesis}. The average grain size of the spherical grains in
the as-grown films was estimated from the SEM micrograph (fig 2a) and 
were found to be $\sim$430nm. The average grain size of this species is
about 140nm in the sample heated to 3hrs. Similar estimation of grain size
in the thicker sample (d=4000\AA, see fig 3) gives the average spherical 
grain's size
as $\sim$220nm. Much variation was not seen in grain size in the 4000\AA\,
samples on heating upto 3 hours. No systematic
variation in grain size was found with sample thickness.

\begin{figure}[h]
\begin{center}
\epsfig{file=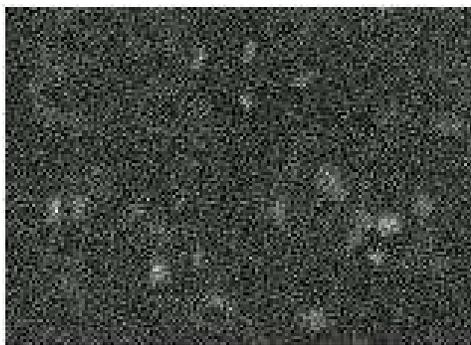,width=2.5in}
\vskip -0.75cm
\caption{SEM micro-graphs of an as-grown sample of film thickness 4000\AA.}
\vskip -2cm
\end{center}
\label{fig:1xa}
\end{figure}

A systematic study of the variation of grain size with heating time could
not be done as samples heated for time greater than 3hrs showed no features
on the surface. While one may infer ${\rm CdI_2}$ to be absent in these
samples, as can be seen from the next section, their existence in the film is 
beyond doubt. One can safely conclude, that the grains migrate/ diffuse into
the film's thickness away from the surface. It is difficult to comment on the 
grain size of the second species of grains
due to it's asymmetrical shape. While the average width of these spindle's
was found to be 140nm, the length of these grains varied from 400nm to 
${\rm 1\mu m}$. Spindle shaped nanocrystals are common and are characterized
by the aspect ratio (length/width). The aspect ratio of the second
species of grains was found to lie between 2.8 to 7.0 in the sample heated for
3 hours.

\subsection{Optical properties of the samples}

Visual examination of the films showed that it had a yellow to reddish-brown
color depending on the thickness of the film. This is in keeping with the
color of ${\rm Sb_2S_3}$ films of the corresponding thickness. On heating,
the color of the films became darker with preceivable change in the film
texture. We have studied the optical absorption of all the samples. 
Typical absorption spectrum of the samples is shown in fig(4) for the 
purpose of illustration. As expected the thicker film (d=4000\AA, labeled
'a') absorbs more strongly than the thinner film (sample labeled 'b'). Both
${\rm Sb_2S_3}$ and ${\rm CdI_2}$ in film state do not strongly absorb in
the wavelength region of 600-900nm and show characteristic fringes (maxima/ 
minima) in the same region. However, in these samples where ${\rm CdI_2}$ is
embedded in ${\rm Sb_2S_3}$ films, very strong absorption peaks are seen at
$\sim$710nm and $\sim$820nm. These peaks are contributed by surface
plasmons. 

\begin{figure}[h]
\begin{center}
\epsfig{file=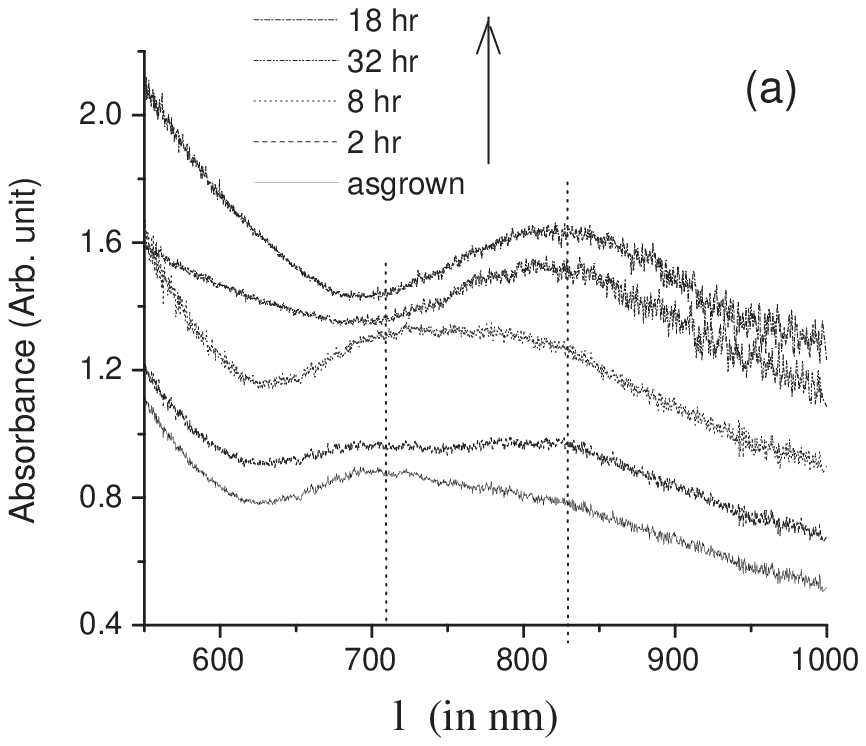,width=3.25in}
\hfil
\epsfig{file=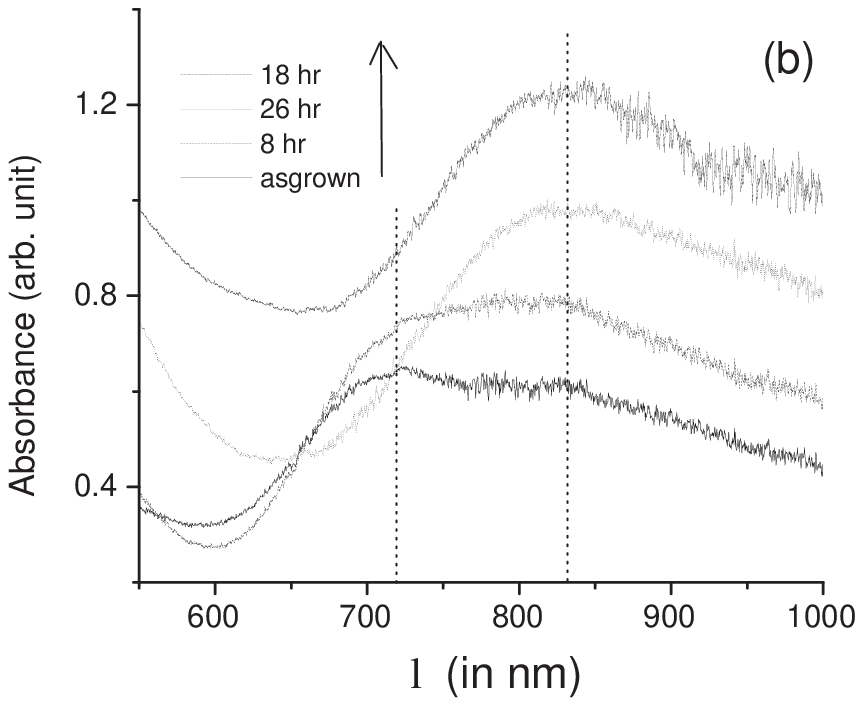, width=3.25in}
\vskip -1cm
\caption{The absorption spectra for (a) 4000\AA\, and (b) 2500\AA\, thick films
with samples heated for various durations. }
\vskip -2cm
\end{center}
\label{fig:1y}
\end{figure}

One can immediately ascribe the two peaks present in the UV-visible spectra to 
the two type of grains (spherical and spindle shaped) present in the film.
Two factors help in associating a peak with each type of grain. Trivially, one 
can immediately correlate these spectra with the results of the SEM studies and 
identify the peak around 710nm with the spherical grains and the peak of 830nm 
to the spindle shaped grains. This inference was extrapolated from the 
variation seen in size with heating till three hours. 
The morphological studies showed that with 
heating the spherical grain's number decreased while number of the spindle 
shaped grains increased. Recollect, it was concluded that on 
heating beyond three hours the 
grains migrated within the film's thickness. Since SEM only 
investigates the sample's surface, one can not conclusively justify this based 
on the morphological studies, however, since the optical studies were done in
transmission mode the response is due to the film as a whole. Thus,
even though grains were not visible in SEM micrographs, the absorption
spectra show spherical grains to be present in the sample even after heating
for eight hours. However, the diminishing peak at 710nm and increasing peak
at 830nm suggests the spherical grains convert into spindle grains,
thereby increasing the population of that grain.

\par Another reason to associate the 830nm peak to the spindle shaped
nano-crystals is due to the nature of the peak itself. Observe the peak's
spread out to the longer wavelength. This results from the fact that the
nanoparticles are spindle shaped having length (b) different from the width
(a) or in other words an non-unity ($\neq$ 1) aspect ratio (b/a). The spread
into longer wavelength becomes greater for large aspect ratio. 

\par As can be
seen the optical properties (SPR) of nanocrystals are also effected by the
particle shape and size. Usually, the decrease in grain size is not
justified by the decreasing peak size but by a blue-shift in the peak
position\cite{blue1,blue2}. This is due to the size dependence of the band
gap in the quantum regime. To establish the peak positions, the broad 
peaks of the UV-visible 
spectra were deconvoluted into two Gaussian using the software "Origin 6.0". 
The variation in peak position with
heating time is shown in fig(5). The points of these graphs are
quite scattered and reflects the difficulty in uniquely identifying the peak
position using the software due to the lack of sharp shoulders in the
spectras. While there is a definite blue shift in the 710nm peaks, the peak
positions of the 830nm increases (red-shifts) with heating. Converse to the 
blue shift which show decrease in nanocrystal's size, the
red shift suggests an increase in nanocrsytal size.

\begin{figure}[h]
\begin{center}
\epsfig{file=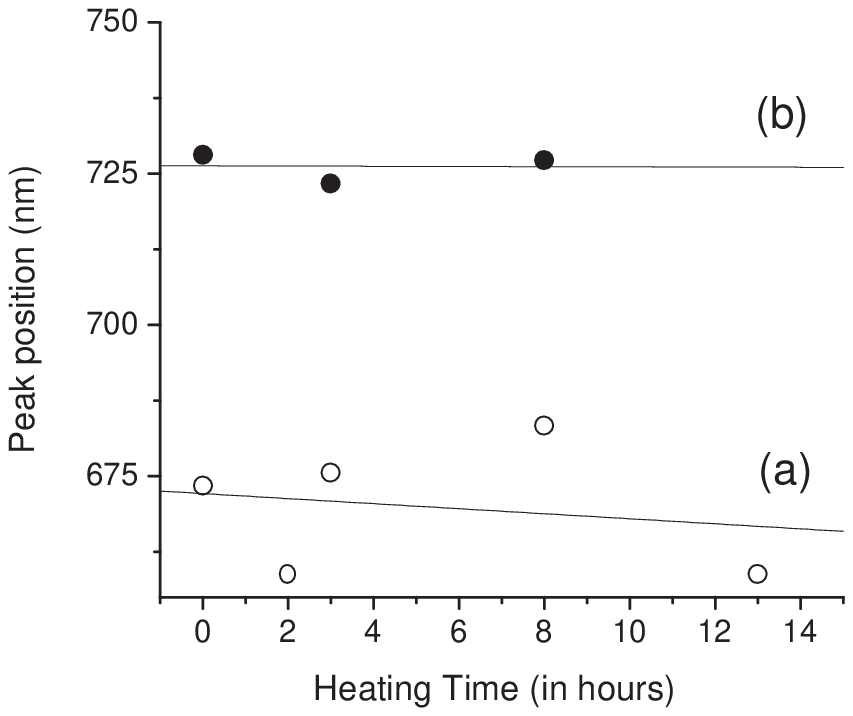,width=3.25in}
\hfil
\epsfig{file=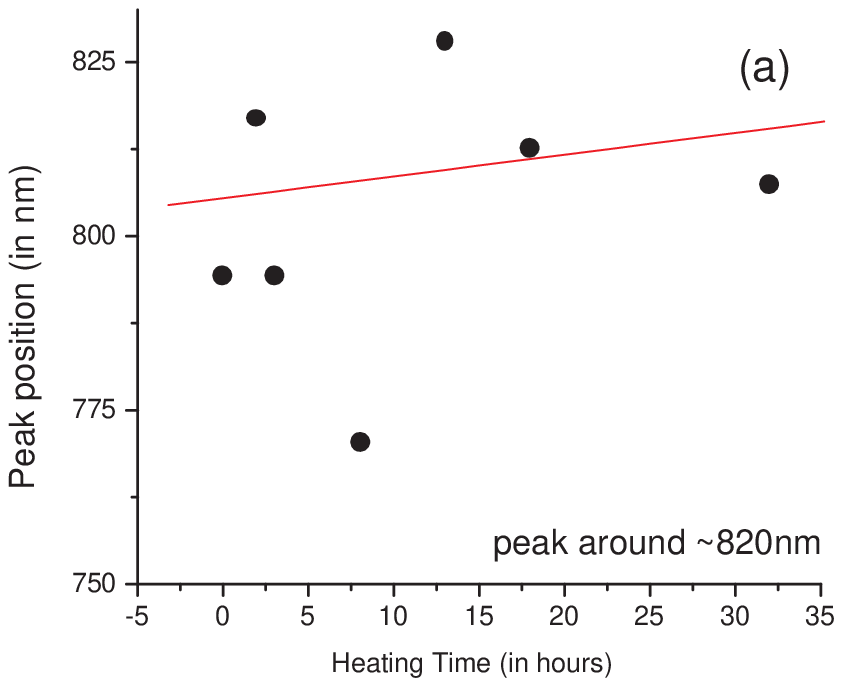, width=3.25in}
\vskip -1cm
\caption{The variation of the peak position with heating time for the peaks
around 710nm and 820nm respectively. The labels (a) and (b) indicate the
sample's thickness as 4000\AA\, and 2500\AA\, respectively.}
\vskip -2cm
\end{center}
\label{fig:14}
\end{figure}

Qualitative estimation of the grain size can be made using the UV-visible
spectra. It has been extensively used to determine the spherical metal 
nanocrystals size/diameter using the equation
\begin{eqnarray}
<D> = {hv_F \over 2\pi \Delta E_{1/2}}\nonumber
\end{eqnarray}
where h is the Planck's constant, ${\rm \Delta E_{1/2}}$ is the full-width at 
half-maximum of the absorption peak and ${\rm v_F}$ is the Fermi velocity of 
electrons. Liang-Fu Lou\cite{wsbook} gives the value of Fermi velocity to be around ${\rm 1
\times 10^8m/s}$. Using this formula, the grain size was calculated as a
function of heating duration. The decrease of grain size with heating time
is shown in fig(6). Notice that the grain size thus estimated and that
evaluated from SEM micrographs are in good agreement. 

\begin{figure}[h]
\begin{center}
\epsfig{file=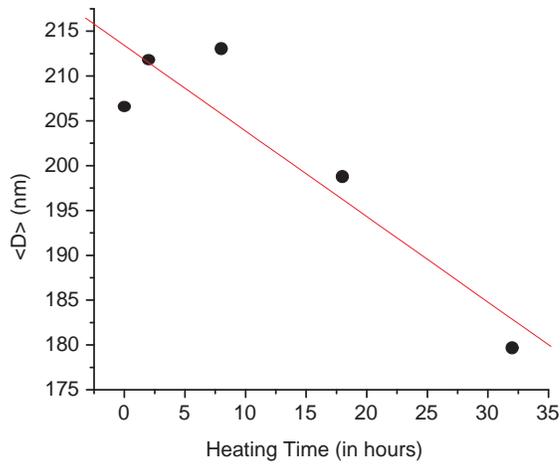,width=3.25in}
\vskip -1cm
\caption{The variation of grain size with heating time for film (a),
thickness 4000\AA. The drawn line is only for visual aid.}
\vskip -2cm
\end{center}
\label{fig:14b}
\end{figure}

\par It would be worthwhile to recollect that ${\rm CdI_2}$ films grown at higher 
deposition rates (2-5nm/sec) lead to films which crystallizes with nH structures 
(${\rm n >4}$). On heating these structures convert into the stable 4H state. 
The diminishing 710nm peak and growing 830nm peak may relate to nH structures 
reducing to 4H. Since the dielectric constant of ${\rm CdI_2}$ films with nH 
structure lies between 60-190 (all possible structures) and that of the stable 
4H structured films is reported to be $\sim$15-30, one may conclude the two
well resolved SPR peaks are due to the diverge dielectric constants of the
nanocrystals.

\section{Conclusion}

Two peaks were found in the UV-visible absorption spectra of ${\rm Sb_2S_3}$ 
films which had ${\rm CdI_2}$ nanocrystals embedded in them. The peaks are
associated with two morphological grains each corresponding to a type of 
crystalline states of ${\rm CdI_2}$ marked by different dielectric
constants. 
The results of this experiments show that as the disparity between the
material acting as the host and the material whose nanocrystals are suspended 
in it increases, there is a distinct blue-shift in the surface plasmon
resonance (SPR).

\section*{Acknowledgement}
\par The author would like to express his sincere gratitude to Dr. A. G. 
Vedeshwar, Department of Physics and Astrophysics, University of Delhi for
allowing to use the facilities of his lab and Dr. N. C. Mehra, University
Science and Instrumentation Center (USIC), University of Delhi, for 
carrying out the SEM analysis.


\end{document}